\documentclass[twocolumn,showpacs,preprintnumbers]{revtex4}%
\usepackage{amssymb}
\usepackage{amsmath}
\usepackage{dcolumn}
\usepackage{bm}
\usepackage{xcolor}
\usepackage[normalem]{ulem}
\usepackage{amsfonts}
\usepackage{graphicx}%
\providecommand{\U}[1]{\protect\rule{.1in}{.1in}}
\providecommand{\U}[1]{\protect\rule{.1in}{.1in}}
\def\showal{1}
\newcommand{\al}[1]{\ifthenelse{\showal=1}{\textcolor{orange}{[[#1]]}}{}}

\newcommand{\eb}[1]{\ifthenelse{\showal=1}{\textcolor{cyan}{[[#1]]}}{}}
\begin{document}
\title{Stability of the Grabert master equation}
\author{Eyal Buks}
\affiliation{Andrew and Erna Viterbi Department of Electrical Engineering, Technion, Haifa
32000, Israel}
\author{Dvir Schwartz}
\affiliation{Andrew and Erna Viterbi Department of Electrical Engineering, Technion, Haifa
32000, Israel}
\date{\today }

\begin{abstract}
We study the dynamics of a quantum system having Hilbert space of finite
dimension $d_{\mathrm{H}}$. Instabilities are possible provided that the
master equation governing the system's dynamics contain nonlinear terms. Here
we consider the nonlinear master equation derived by Grabert. The dynamics
near a fixed point is analyzed by using the method of linearization, and by
evaluating the eigenvalues of the Jacobian matrix. We find that all these
eigenvalues are non-negative, and conclude that the fixed point is stable.
This finding raises the question: under what conditions instability is
possible in a quantum system having finite $d_{\mathrm{H}}$?

\end{abstract}
\maketitle





Consider a given closed quantum system having Hilbert space of finite
dimension $d_{\mathrm{H}}$, whose master equation, which governs the time
evolution of the reduced density matrix $\rho$,\ can be expressed as
$\mathrm{d}\rho/\mathrm{d}t=$ $\Theta\left(  \rho\right)  =\Theta_{\mathrm{u}%
}\left(  \rho\right)  -\Theta_{\mathrm{d}}\left(  \rho\right)  $. The first
term, which is given by $\Theta_{\mathrm{u}}\left(  \rho\right)  =\left(
i/\hbar\right)  \left[  \rho,\mathcal{H}\right]  $, where $\mathcal{H}%
=\mathcal{H}^{\dag}$\ is the Hamiltonian of the closed system, represents
unitary evolution, and the second one $\Theta_{\mathrm{d}}\left(  \rho\right)
$ represents the effect of coupling between the closed system and its
environment. While it is commonly assumed that both the unitary term
$\Theta_{\mathrm{u}}\left(  \rho\right)  $ and the damping term $\Theta
_{\mathrm{d}}\left(  \rho\right)  $ are linear in $\rho$
\cite{Fernengel_385701,Lindblad_119}, in some cases the master equation can
become nonlinear. Two types of nonlinearity are considered below
\cite{Yukalov_9232,Prataviera_01}. For the first one, which is henceforth
referred to as unitary nonlinearity, the unitary term $\Theta_{\mathrm{u}%
}\left(  \rho\right)  $ is replaced by a nonlinear term. In most cases,
unitary nonlinearity originates from either the mean field approximation
\cite{breuer2002theory,Drossel_217,Hicke_024401,Levi_053516}, or from a
transformation mapping the Hilbert space of finite dimension $d_{\mathrm{H}}$
into a space having infinite dimensionality (e.g. the Holstein-Primakoff
transformation \cite{Holstein_1098}, which can yield a parametric instability
in ferromagnetic resonators \cite{Schlomann_S386}). Here, we consider the
second type, which is henceforth referred to as damping nonlinearity, and
focus on the master equation that was proposed by Grabert \cite{Grabert_161},
which has a damping term $\Theta_{\mathrm{d}}\left(  \rho\right)  $ nonlinear
in $\rho$.

Grabert has shown that the invalidity of the quantum regression hypothesis
gives rise to damping nonlinearity \cite{Grabert_161}. The
nonlinear term added to the master equation ensures that the purity
$\operatorname{Tr}\rho^{2}$ does not exceed unity
\cite{Ottinger_052119,Ottinger_10006}, and that entropy is generated at a
non-negative rate, as is expected from the second law of thermodynamics
\cite{Taj_062128}. Note, however, that under appropriate conditions, nonlinear
dynamics may allow for faster than light signaling \cite{Bassi_055027}.

The Grabert master equation (GME) has a fixed point given by%
\begin{equation}
\rho_{0}=\frac{e^{-\beta\mathcal{H}}}{\operatorname{Tr}\left(  e^{-\beta
\mathcal{H}}\right)  }\;, \label{rho TE}%
\end{equation}
where $\beta=1/\left(  k_{\mathrm{B}}T\right)  $ is the inverse of the thermal
energy \cite{Grabert_161}. At the fixed point $\rho_{0}$ the system is in
thermal equilibrium having Boltzmann distribution.

Here we explore the stability of this fixed point $\rho_{0}$ for the case
where the Hamiltonian $\mathcal{H}$ of the closed system is time-independent.
In a basis of energy eigenstates of a time-independent Hamiltonian both
matrices $\mathcal{H}=\operatorname{diag}\left(  E_{1},E_{2},\cdots
,E_{d_{\mathrm{H}}}\right)  $ and $\rho_{0}=\operatorname{diag}\left(
\rho_{1},\rho_{2},\cdots,\rho_{d_{\mathrm{H}}}\right)  $ are diagonal, where
$\rho_{n}=e^{-\beta E_{n}}/\operatorname{Tr}\left(  e^{-\beta\mathcal{H}%
}\right)  $ [see Eq. (\ref{rho TE})].

For the case of thermal equilibrium, one may argue that the
stability of $\rho_{0}$ is obvious. However, the stability of a driven system
is anything but obvious. Note that in many cases the rotating wave
approximation (RWA) is employed in order to model the dynamics of a given
system under external driving, using a transformation into a rotating frame,
in which the Hamiltonian becomes time-independent in the RWA. Thus, our
conclusion, that the fixed point $\rho_{0}$ is stable for any time independent
Hermitian $\mathcal{H}$, can be extended beyond the limits of thermal equilibrium.

The GME for the reduced density matrix $\rho$ can be expressed as
\cite{Ottinger_052119}%
\begin{equation}
\frac{\mathrm{d}\rho}{\mathrm{d}t}=\Theta\left(  \rho\right)  =\Theta
_{\mathrm{u}}\left(  \rho\right)  -\Theta_{\mathrm{d}}\left(  \rho\right)  \;,
\label{TME}%
\end{equation}
where the damping term is given by $\Theta_{\mathrm{d}}\left(  \rho\right)
=\Theta_{\mathrm{A}}\left(  \rho\right)  +\Theta_{\mathrm{B}}\left(
\rho\right)  $, where $\Theta_{\mathrm{A}}\left(  \rho\right)  $, which is
given by $\Theta_{\mathrm{A}}\left(  \rho\right)  =\gamma_{\mathrm{E}}\left[
Q,\left[  Q,\rho\right]  \right]  $, is linear in $\rho$, and $\Theta
_{\mathrm{B}}\left(  \rho\right)  $, which is given by $\Theta_{\mathrm{B}%
}\left(  \rho\right)  =\beta\gamma_{\mathrm{E}}\left[  Q,\left[
Q,\mathcal{H}\right]  _{\rho}\right]  $ is nonlinear. The constant
$\gamma_{\mathrm{E}}>0$ is a damping rate, the Hermitian operator $Q^{\dag}=Q$
describes the interaction between the quantum system and its environment, and%
\begin{equation}
A_{\rho}=\int_{0}^{1}\mathrm{d}\eta\;\rho^{\eta}A\rho^{1-\eta}\;.
\label{A_rho}%
\end{equation}

Alternatively, the damping term $\Theta_{\mathrm{d}}\left(  \rho\right)  $ can
be expressed as $\Theta_{\mathrm{d}}\left(  \rho\right)  =\beta\gamma
_{\mathrm{E}}\left[  Q,\left[  Q,\mathcal{U}_{\mathrm{H}}\right]  _{\rho
}\right]  $, where $\mathcal{U}_{\mathrm{H}}=\mathcal{H}+\beta^{-1}\log\rho$
is the Helmholtz free energy operator \cite{Grabert_161}. According to the
master equation (\ref{TME}), the time evolution of the Helmholtz free energy
$\left\langle \mathcal{U}_{\mathrm{H}}\right\rangle =\operatorname{Tr}\left(
\mathcal{U}_{\mathrm{H}}\rho\right)  $ is governed by%
\begin{equation}
\frac{\mathrm{d}\left\langle \mathcal{U}_{\mathrm{H}}\right\rangle
}{\mathrm{d}t}=-\beta\gamma_{\mathrm{E}}\operatorname{Tr}\left(
\mathcal{C}_{\rho}\mathcal{C}\right)  \;, \label{HFE eom}%
\end{equation}
where $\mathcal{C}=i\left[  Q,\mathcal{U}_{\mathrm{H}}\right]  $, and thus
$\mathrm{d}\left\langle \mathcal{U}_{\mathrm{H}}\right\rangle /\mathrm{d}%
t\leq0$ (since $\mathcal{C}^{\dag}=\mathcal{C}$)
\cite{Ottinger_052119,kubo2012statistical}, i.e. the Helmholtz free energy
$\left\langle \mathcal{U}_{\mathrm{H}}\right\rangle $ is a monotonically
decreasing function of time.

Note that the operator $\mathcal{C}$ vanishes at the fixed point $\rho_{0}$
given by Eq. (\ref{rho TE}). Alternatively, the Kubo's identity [given by Eq.
(4.2.17) of \cite{kubo2012statistical}] can be used to show that $\rho_{0}$ is
a fixed point \cite{Grabert_161}. For some cases the existence of a limit
cycle (i.e. periodic) solution for the GME (\ref{TME}) can be ruled out using
Eq. (\ref{HFE eom}). Along such a solution the condition $\mathcal{C}=0$ must
be satisfied [since $\operatorname{Tr}\left(  \mathcal{C}_{\rho}%
\mathcal{C}\right)  =0$ implies that $\mathcal{C}=0$ when $\operatorname{Tr}%
\rho^{2}<1$]. Hence, when $\rho=\rho_{0}$ is a unique solution of
$\mathcal{C}=0$, a limit cycle solution can be ruled out.

A linear master equation can be derived by replacing the nonlinear term
$\Theta_{\mathrm{B}}\left(  \rho\right)  $ by the term $\left(  \beta^{\prime
}/\hbar\right)  \gamma_{\mathrm{E}}\left[  Q,\left[  Q,\mathcal{H}\right]
\right]  $, where $\beta^{\prime}>0$. It was shown in Ref. \cite{Spohn_33}
(see also appendix B of Ref. \cite{Levi_053516}) that such a linear master
equation is stable provided that $\gamma_{\mathrm{E}}>0$. Below we analyze the
stability of the nonlinear GME (\ref{TME}).

The stability of the fixed point $\rho_{0}$ of the master equation (\ref{TME})
is explored by the method of linearization applied to the nonlinear term
$\Theta_{\mathrm{B}}\left(  \rho\right)  $. In the vicinity of $\rho
_{0}=\operatorname{diag}\left(  \rho_{1},\rho_{2},\cdots,\rho_{d_{\mathrm{H}}%
}\right)  $ the density matrix $\rho$ is expressed as $\rho=\rho_{0}%
+\epsilon\mathcal{V}$, where $\epsilon$ is a real small parameter. Let $u\rho
u^{\dag}=\rho_{\mathrm{d}}=\operatorname{diag}\left(  \rho_{1}^{\prime}%
,\rho_{2}^{\prime},\cdots,\rho_{d_{\mathrm{H}}}^{\prime}\right)  $ be
diagonal, where $u$ is unitary, i.e. $u^{\dag}u=1$. With the help of
time-independent perturbation theory one finds that the eigenvalues $\rho
_{n}^{\prime}$ of $\rho$\ are given by%
\begin{equation}
\rho_{n}^{\prime}=\rho_{n}+\epsilon\left(  n\right\vert \mathcal{V}\left\vert
n\right)  +O\left(  \epsilon^{2}\right)  \;,\label{rho_n'}%
\end{equation}
and the unitary transformation $u$ that diagonalizes $\rho$ is given by%
\begin{align}
u &  =\sum_{n}\left(  \left\vert n\right)  +\sum_{k\neq n}\frac{\epsilon
\left(  k\right\vert \mathcal{V}\left\vert n\right)  }{\rho_{n}-\rho_{k}%
}\left\vert k\right)  \right)  \left(  n\right\vert +O\left(  \epsilon
^{2}\right)  \;,\nonumber\\
&
\end{align}
or%
\begin{equation}
u=1-i\epsilon F+O\left(  \epsilon^{2}\right)  \;,\label{u=1-i*F}%
\end{equation}
where the Hermitian matrix $F$ is given by%
\begin{equation}
F=\sum_{k\neq l}\frac{i\left(  k\right\vert \mathcal{V}\left\vert l\right)
}{\rho_{l}-\rho_{k}}\left\vert k\right)  \left(  l\right\vert \;,\label{F=}%
\end{equation}
$\left(  k\right\vert \mathcal{V}\left\vert l\right)  =\mathcal{V}_{kl}$ is
the ($k$'th raw - $l$'th column) matrix element of $\mathcal{V}$, and
$\left\vert k\right)  \left(  l\right\vert $ denotes a $d_{\mathrm{H}}\times
d_{\mathrm{H}}$ matrix having entry $1$ in the ($k$'th raw - $l$'th column),
and entry $0$ elsewhere.

Using the identity \cite{Ottinger_052119}%
\begin{equation}
\int_{0}^{1}x^{\eta}y^{1-\eta}\mathrm{d}\eta=\mathcal{F}\left(  x,y\right)
\;,
\end{equation}
where%
\begin{equation}
\mathcal{F}\left(  x,y\right)  =\frac{x-y}{\log x-\log y}\;, \label{F(x,y) V1}%
\end{equation}
one finds that (recall that $\rho_{\mathrm{d}}$ is diagonal)%
\begin{equation}
\int_{0}^{1}\mathrm{d}\eta\;\rho_{\mathrm{d}}^{\eta}A\rho_{\mathrm{d}}%
^{1-\eta}=\mathcal{F}^{\prime}\circ A\;, \label{A_rho Pi}%
\end{equation}
where $\circ$ denotes the Hadamard matrix multiplication (element by element
matrix multiplication), and where the matrix elements of $\mathcal{F}^{\prime
}$ are given by $\mathcal{F}_{nm}^{\prime}=\mathcal{F}\left(  \rho_{n}%
^{\prime},\rho_{m}^{\prime}\right)  $. Note that $\mathcal{F}_{nm}^{\prime
}=\mathcal{F}_{nm}+O\left(  \epsilon\right)  $, where $\mathcal{F}%
_{nm}=\mathcal{F}\left(  \rho_{n},\rho_{m}\right)  $ [see Eq. (\ref{rho_n'})],
hence, the following holds [see Eqs. (\ref{A_rho}) and (\ref{u=1-i*F}) and
note that $uAu^{\dag}=A+i\epsilon\left[  A,F\right]  +O\left(  \epsilon
^{2}\right)  $]%
\begin{equation}
A_{\rho}=\mathcal{F}^{\prime}\circ A+i\epsilon\left[  F,\mathcal{F}\circ
A\right]  +i\epsilon\mathcal{F}\circ\left[  A,F\right]  +O\left(  \epsilon
^{2}\right)  \;, \label{A_rho PT}%
\end{equation}
where $\mathcal{F}^{\prime}=\mathcal{F}+\epsilon\left(  \mathrm{d}%
\mathcal{F}/\mathrm{d}\epsilon\right)  +O\left(  \epsilon^{2}\right)  $.

The following holds [see Eq. (\ref{F(x,y) V1})]%
\begin{equation}
\mathcal{F}\left(  x,y\right)  =\frac{x+y}{2}f_{\mathrm{D}}\left(  \frac
{x-y}{x+y}\right)  \;, \label{F(x,y) V2}%
\end{equation}
where the function $f_{\mathrm{D}}\left(  \eta\right)  $ is given by%
\begin{equation}
f_{\mathrm{D}}\left(  \eta\right)  =\frac{2\eta}{\log\frac{1+\eta}{1-\eta}%
}=\frac{\eta}{\tanh^{-1}\eta}\;. \label{f_D}%
\end{equation}
The function $f_{\mathrm{D}}$ is symmetric, i.e. $f_{\mathrm{D}}\left(
-\eta\right)  =f_{\mathrm{D}}\left(  \eta\right)  $, and the following holds
$f_{\mathrm{D}}\left(  0\right)  =1$ and $f_{\mathrm{D}}\left(  \pm1\right)
=0$. With the help of Eqs. (\ref{rho_n'}) and (\ref{F(x,y) V2}) one finds that
the matrix $\mathrm{d}\mathcal{F}/\mathrm{d}\epsilon\ $is real, symmetric, and
the following holds (no summation due to repeated indices $n$ and $m$)%
\begin{equation}
\left(  \frac{\mathrm{d}\mathcal{F}}{\mathrm{d}\epsilon}\right)  _{nm}%
=\frac{\mathrm{d}\alpha_{nm}}{\mathrm{d}\epsilon}F_{nm}+\alpha_{nm}%
\frac{\mathrm{d}\eta_{nm}}{\mathrm{d}\epsilon}F_{nm}^{\prime}\;,
\label{(dF/d epsilon)}%
\end{equation}
where $\alpha_{nm}=\left(  \rho_{n}+\rho_{m}\right)  /2$, $\eta_{nm}=\left(
\rho_{n}-\rho_{m}\right)  /\left(  \rho_{n}+\rho_{m}\right)  $, $F_{nm}%
=f_{\mathrm{D}}\left(  \eta_{nm}\right)  $, and where $F_{nm}^{\prime
}=f_{\mathrm{D}}^{\prime}\left(  \eta_{nm}\right)  $. Moreover,
$\operatorname{Tr}\left(  \mathrm{d}\mathcal{F}/\mathrm{d}\epsilon\right)  =0$
(note that $F_{nn}=1$ and $F_{nn}^{\prime}=0$).

The $d_{\mathrm{H}}^{2}-1$ Hermitian and trace-less $d_{\mathrm{H}}\times
d_{\mathrm{H}}$ generalized Gell-Mann matrices $\lambda_{n}$, which span the
SU($d_{\mathrm{H}}$) Lie algebra, satisfy the orthogonality relation%
\begin{equation}
\frac{\operatorname{Tr}\left(  \lambda_{a}\lambda_{b}\right)  }{2}=\delta
_{ab}\;. \label{Tr(lambda_a*lambda_b)=}%
\end{equation}
For the case $d_{\mathrm{H}}=2$ ($d_{\mathrm{H}}=3$) the matrices are called
Pauli (Gell-Mann) matrices. The set $\left\{  \lambda_{a}\right\}  $ of
$d_{\mathrm{H}}^{2}-1$ matrices can be divided into three subsets. The subset
$\left\{  \lambda_{\mathrm{X},\left(  n,m\right)  }\right\}  $ contains
$d_{\mathrm{H}}\left(  d_{\mathrm{H}}-1\right)  /2$ matrices given by
$\lambda_{\mathrm{X},\left(  n,m\right)  }=\left\vert n\right)  \left(
m\right\vert +\left\vert m\right)  \left(  n\right\vert $, and the subset
$\left\{  \lambda_{\mathrm{Y},\left(  n,m\right)  }\right\}  $ contains
$d_{\mathrm{H}}\left(  d_{\mathrm{H}}-1\right)  /2$ matrices given by
$\lambda_{\mathrm{Y},\left(  n,m\right)  }=-i\left\vert n\right)  \left(
m\right\vert +i\left\vert m\right)  \left(  n\right\vert $, where $1\leq
m<n\leq d_{\mathrm{H}}$. The subset $\left\{  \lambda_{\mathrm{Z},l}\right\}
$ contains $d_{\mathrm{H}}-1$ diagonal matrices given by%
\begin{equation}
\lambda_{\mathrm{Z},l}=\sqrt{\frac{2}{l\left(  l+1\right)  }}\left(
-l\left\vert l+1\right)  \left(  l+1\right\vert +\sum_{j=1}^{l}\left\vert
j\right)  \left(  j\right\vert \right)  \;, \label{lambda Z}%
\end{equation}
where $1\leq l\leq d_{\mathrm{H}}-1$.

It is convenient to express the perturbation $\epsilon\mathcal{V}=\rho
-\rho_{0}$ as $\epsilon\mathcal{V}=\bar{\kappa}\cdot\bar{\lambda}$, where
$\bar{\kappa}=\left(  \kappa_{1},\kappa_{2},\cdots,\kappa_{d_{\mathrm{H}}%
^{2}-1}\right)  $ and $\bar{\lambda}=\left(  \lambda_{1},\lambda_{2}%
,\cdots,\lambda_{d_{\mathrm{H}}^{2}-1}\right)  $. In this notation the GME
(\ref{TME}) becomes (repeated index implies summation)%
\begin{equation}
\frac{\mathrm{d}\kappa_{b}}{\mathrm{d}t}\lambda_{b}=\Theta\left(  \rho
_{0}+\kappa_{b}\lambda_{b}\right)  \;,
\end{equation}
or [see Eq. (\ref{Tr(lambda_a*lambda_b)=})]%
\begin{equation}
\frac{\mathrm{d}\kappa_{a}}{\mathrm{d}t}=\frac{1}{2}\operatorname{Tr}\left(
\Theta\left(  \rho_{0}+\kappa_{b}\lambda_{b}\right)  \lambda_{a}\right)  \;.
\end{equation}
To first order in $\bar{\kappa}$%
\begin{equation}
\frac{\mathrm{d}\kappa_{a}}{\mathrm{d}t}=\frac{1}{2}\operatorname{Tr}\left(
\frac{\partial\Theta}{\partial\kappa_{b}}\lambda_{a}\kappa_{b}\right)  \;,
\end{equation}
or in a vector form%
\begin{equation}
\frac{\mathrm{d}\bar{\kappa}}{\mathrm{d}t}=J\bar{\kappa}\;, \label{LME J}%
\end{equation}
where the Jacobian matrix $J$ is given by $J=J_{\mathrm{u}}-J_{\mathrm{A}%
}-J_{\mathrm{B}}$, and where%
\begin{equation}
J_{\Sigma}=\frac{1}{2}\operatorname{Tr}\left(  \frac{\partial\Theta_{\Sigma}%
}{\partial\kappa_{b}}\lambda_{a}\right)  \;, \label{J_Sigma}%
\end{equation}
with $\Sigma\in\left\{  \mathrm{u},\mathrm{A},\mathrm{B}\right\}  $.

The system's stability depends on the set of eigenvalues of the Jacobian
matrix $J$, which is denoted by $\mathcal{S}$. The system is stable provided
that $\operatorname{real}\left(  \xi\right)  <0$ for any $\xi\in\mathcal{S}$.
It was shown in appendix B of Ref. \cite{Levi_053516} that such a system is
stable provided that $J_{\mathrm{u}}$, $J_{\mathrm{A}}$ and $J_{\mathrm{B}}$
are all real, $J_{\mathrm{u}}$ is antisymmetric, all diagonal elements of
$J_{\mathrm{A}}+J_{\mathrm{B}}$ are positive, and $d_{\mathrm{H}}$ is finite.
Properties of the matrices $J_{\mathrm{u}}$, $J_{\mathrm{A}}$ and
$J_{\mathrm{B}}$ are analyzed below.

The matrix $J_{\mathrm{u}}$, which governs the unitary evolution, is given by
[recall the trace identity $\operatorname{Tr}\left(  XY\right)
=\operatorname{Tr}\left(  YX\right)  $]%
\begin{align}
J_{\mathrm{u}}  &  =\frac{i}{2\hbar}\operatorname{Tr}\left(  \left[
\lambda_{b},\mathcal{H}\right]  \lambda_{a}\right) \nonumber\\
&  =\frac{i}{2\hbar}\operatorname{Tr}\left(  \mathcal{H}\left[  \lambda
_{a},\lambda_{b}\right]  \right)  \;,\nonumber\\
&  \label{J_u}%
\end{align}
hence $J_{\mathrm{u}}$ is \textit{real} and \textit{antisymmetric} provided
that $\mathcal{H}$ is Hermitian (note that $i\left[  \lambda_{b},\lambda
_{a}\right]  $ is Hermitian).

The matrix $J_{\mathrm{A}}$ is given by%
\begin{align}
J_{\mathrm{A}}  &  =\frac{\gamma_{\mathrm{E}}}{2}\operatorname{Tr}\left(
\left[  Q,\left[  Q,\lambda_{b}\right]  \right]  \lambda_{a}\right)
\nonumber\\
&  =\frac{\gamma_{\mathrm{E}}}{2}\operatorname{Tr}\left(  -\left[
Q,\lambda_{b}\right]  \left[  Q,\lambda_{a}\right]  \right)  \;.\nonumber\\
&  \label{J_A}%
\end{align}
Both matrices $i\left[  Q,\lambda_{a}\right]  $ and $i\left[  Q,\lambda
_{b}\right]  $ are Hermitian, provided that $Q$ is Hermitian, hence
$J_{\mathrm{A}}$ is \textit{real} (recall that $\gamma_{\mathrm{E}}$ is
positive). The \textit{diagonal} elements of $J_{\mathrm{A}}$ are
\textit{positive} since $-\left[  Q,\lambda_{b}\right]  \left[  Q,\lambda
_{a}\right]  $ is positive-definite for the case $a=b$.

The diagonal elements of the matrix $J_{\mathrm{B}}$ cab be evaluated using
the linearization of the term $A_{\rho}$ given by Eq. (\ref{A_rho PT}). For
the case where the perturbation $\mathcal{V}=\left(  \rho-\rho_{0}\right)
/\epsilon$ is a generalized Gell-Mann matrix, i.e. $\mathcal{V}\in\left\{
\lambda_{a}\right\}  $, the following holds [see Eq. (\ref{F=})]%
\begin{equation}
F=\left\{
\begin{array}
[c]{cc}%
\frac{\lambda_{\mathrm{Y},\left(  n,m\right)  }}{\rho_{n}-\rho_{m}} & \text{if
}\mathcal{V}=\lambda_{\mathrm{X},\left(  n,m\right)  }\\
-\frac{\lambda_{\mathrm{X},\left(  n,m\right)  }}{\rho_{n}-\rho_{m}} &
\text{if }\mathcal{V}=\lambda_{\mathrm{Y},\left(  n,m\right)  }%
\end{array}
\right.  \;, \label{F V}%
\end{equation}
and [see Eq. (\ref{A_rho PT}), and note that, according to Eq. (\ref{rho_n'}),
$\mathcal{F}^{\prime}=\mathcal{F}+O\left(  \epsilon^{2}\right)  $ when all
diagonal elements of the perturbation vanish, e.g. when $\mathcal{V}%
\in\left\{  \lambda_{\mathrm{X},\left(  n,m\right)  }\right\}  \cup\left\{
\lambda_{\mathrm{Y},\left(  n,m\right)  }\right\}  $, and, according to Eqs.
(\ref{u=1-i*F}) and (\ref{F=}), $u=1+O\left(  \epsilon^{2}\right)  $ when the
perturbation is diagonal, e.g. when $\mathcal{V}\in\left\{  \lambda
_{\mathrm{Z},l}\right\}  $]%
\begin{equation}
\frac{\mathrm{d}A_{\rho}}{\mathrm{d}\epsilon}=\left\{
\begin{array}
[c]{cc}%
\frac{\left[  \mathcal{F}\circ A,\lambda_{\mathrm{Y},\left(  n,m\right)
}\right]  -\mathcal{F}\circ\left[  A,\lambda_{\mathrm{Y},\left(  n,m\right)
}\right]  }{i\left(  \rho_{n}-\rho_{m}\right)  } & \text{if }\mathcal{V}%
=\lambda_{\mathrm{X},\left(  n,m\right)  }\\
\frac{\left[  \mathcal{F}\circ A,\lambda_{\mathrm{X},\left(  n,m\right)
}\right]  -\mathcal{F}\circ\left[  A,\lambda_{\mathrm{X},\left(  n,m\right)
}\right]  }{\left(  -i\right)  \left(  \rho_{n}-\rho_{m}\right)  } & \text{if
}\mathcal{V}=\lambda_{\mathrm{Y},\left(  n,m\right)  }\\
\frac{\mathrm{d}\mathcal{F}^{\prime}}{\mathrm{d}\epsilon}\circ A & \text{if
}\mathcal{V}=\lambda_{\mathrm{Z},\left(  n,m\right)  }%
\end{array}
\right.  \;. \label{d A_rho / d epsilon V}%
\end{equation}

The diagonal elements of $J_{\mathrm{A}}+J_{\mathrm{B}}$ are evaluated by
using of Eq. (\ref{d A_rho / d epsilon V}) with different values of the
perturbation $\mathcal{V}$.

The diagonal matrix element corresponding to the generalized Gell-Mann matrix
$\lambda_{\mathrm{Z},l}$, which is labeled by $j_{l}$, is given by [see Eqs.
(\ref{J_Sigma}), (\ref{J_A}) and (\ref{d A_rho / d epsilon V})]%
\begin{align}
j_{l}  &  =\frac{\gamma_{\mathrm{E}}}{2}\operatorname{Tr}\left(  -\left[
Q,\lambda_{\mathrm{Z},l}\right]  ^{2}\right) \nonumber\\
&  +\frac{\beta\gamma_{\mathrm{E}}}{2}\operatorname{Tr}\left(  \left[
Q,\frac{\mathrm{d}\mathcal{F}}{\mathrm{d}\epsilon}\circ\left[  Q,\mathcal{H}%
\right]  \right]  \lambda_{\mathrm{Z},l}\right)  \;,\nonumber\\
&  \label{j_l V1}%
\end{align}
where the term $\mathrm{d}\mathcal{F}/\mathrm{d}\epsilon$ is evaluated
according to Eq. (\ref{(dF/d epsilon)}) for the case where the perturbation is
given by $\mathcal{V}=\lambda_{\mathrm{Z},l}$. In terms of the elements of the
diagonal matrix $\lambda_{\mathrm{Z},l}=\operatorname{diag}\left(  \nu_{1}%
,\nu_{2},\cdots,\nu_{d_{\mathrm{H}}}\right)  $ one finds using Eq.
(\ref{rho_n'}) that $\rho_{n}^{\prime}=\rho_{n}+\epsilon\nu_{n}+O\left(
\epsilon^{2}\right)  $, hence $\left(  \mathrm{d}\mathcal{F}/\mathrm{d}%
\epsilon\right)  _{nm}=d_{nm}$, where%
\begin{equation}
d_{nm}=\frac{\nu_{nm}F_{nm}}{2\varkappa_{nm}}\left(  1+\frac{\left(
\varkappa_{nm}-\eta_{nm}\right)  F_{nm}^{\prime}}{F_{nm}}\right)  \;,
\end{equation}
$\nu_{nm}=\nu_{n}-\nu_{m}$ and $\varkappa_{nm}=\left(  \nu_{n}-\nu_{m}\right)
/\left(  \nu_{n}+\nu_{m}\right)  $. The following holds $d_{nm}=d_{mn}$, hence
Eq. (\ref{j_l V1}) yields%
\begin{equation}
j_{l}=\gamma_{\mathrm{E}}\sum_{n<m}\zeta_{nm}\nu_{nm}^{2}\left\vert
q_{nm}\right\vert ^{2}\;, \label{j_l}%
\end{equation}
where $\zeta_{nm}=1+d_{nm}e_{nm}/\nu_{nm}$, $e_{nm}=\beta\left(  E_{n}%
-E_{m}\right)  $, and where $q_{nm}$ are the matrix elements of the operator
$Q$ (recall that it is assumed that $Q^{\dag}=Q$, i.e. $q_{mn}=q_{nm}^{\ast}%
$). With the help of the relation $\eta_{nm}=-\tanh\left(  e_{nm}/2\right)  $
[see Eq. (\ref{rho TE})] one finds that $\zeta_{nm}=\zeta\left(  \eta
_{nm},\varkappa_{nm}\right)  $, where the function $\zeta\left(
\eta,\varkappa\right)  $ is given by [see Eq. (\ref{f_D}) and note that
$1-\left(  1/\left(  1-\eta^{2}\right)  \right)  \left(  \eta/\tanh^{-1}%
\eta\right)  =\eta F^{\prime}\left(  \eta\right)  /F\left(  \eta\right)  $]%
\begin{equation}
\zeta\left(  \eta,\varkappa\right)  =\frac{f_{\mathrm{D}}\left(  \eta\right)
}{1-\eta^{2}}\left(  1-\frac{\eta}{\varkappa}\right)  \;.
\label{zeta(eta,kappa)}%
\end{equation}
The following holds [see Eq. (\ref{lambda Z}), and note that only the cases
for which $v_{nm}\neq0$, i.e. the cases that can contribute to $j_{l}$, are
listed]%
\begin{equation}
-\frac{1}{\varkappa_{nm}}=\left\{
\begin{array}
[c]{cc}%
\frac{l-1}{l+1} & n\leq l\text{ and }m=l+1\\
1 & n\leq l\text{ and }m>l+1\\
1 & n=l+1\text{ and }m>l+1
\end{array}
\right.  \;,
\end{equation}
hence $0\leq\left(  -1/\varkappa\right)  \leq1$ for all terms contributing to
$j_{l}$, hence $\zeta_{nm}\nu_{nm}^{2}\geq0$ for these terms, and consequently
$j_{l}\geq0$.

The diagonal matrix element corresponding to the generalized Gell-Mann matrix
$\lambda_{\mathrm{X},\left(  2,1\right)  }$ ($\lambda_{\mathrm{Y},\left(
2,1\right)  }$) is labelled by $j_{\mathrm{X}}$ ($j_{\mathrm{Y}}$). We show
below that both $j_{\mathrm{X}}$ and $j_{\mathrm{Y}}$ are non-negative. The
proof is applicable for all other diagonal elements, corresponding to all
generalized Gell-Mann matrices $\lambda\in\left\{  \lambda_{\mathrm{X},\left(
n,m\right)  }\right\}  \cup\left\{  \lambda_{\mathrm{Y},\left(  n,m\right)
}\right\}  $ with $\left(  n,m\right)  \neq\left(  2,1\right)  $, since the
ordering of the energy eigenvectors is arbitrary.

With the help of Eqs. (\ref{J_Sigma}), (\ref{J_A}) and
(\ref{d A_rho / d epsilon V}) one finds that [the subscript $\left(
2,1\right)  $ is omitted for brevity]%
\begin{align}
\frac{j_{\mathrm{X}}}{\frac{\gamma_{\mathrm{E}}}{2}}  &  =\operatorname{Tr}%
\left(  -\left[  Q,\lambda_{\mathrm{X}}\right]  \left[  Q,\lambda_{\mathrm{X}%
}\right]  \right) \nonumber\\
&  +\operatorname{Tr}\left(  \beta\left[  Q,\frac{\left[  \mathcal{F}%
\circ\left[  Q,\mathcal{H}\right]  ,\lambda_{\mathrm{Y}}\right]
-\mathcal{F}\circ\left[  \left[  Q,\mathcal{H}\right]  ,\lambda_{\mathrm{Y}%
}\right]  }{i\left(  \rho_{2}-\rho_{1}\right)  }\right]  \lambda_{\mathrm{X}%
}\right)  \;,\nonumber\\
&
\end{align}
and%
\begin{align}
\frac{j_{\mathrm{Y}}}{\frac{\gamma_{\mathrm{E}}}{2}}  &  =\operatorname{Tr}%
\left(  -\left[  Q,\lambda_{\mathrm{Y}}\right]  \left[  Q,\lambda_{\mathrm{Y}%
}\right]  \right) \nonumber\\
&  +\operatorname{Tr}\left(  \beta\left[  Q,\frac{\left[  \mathcal{F}%
\circ\left[  Q,\mathcal{H}\right]  ,\lambda_{\mathrm{X}}\right]
-\mathcal{F}\circ\left[  \left[  Q,\mathcal{H}\right]  ,\lambda_{\mathrm{X}%
}\right]  }{\left(  -i\right)  \left(  \rho_{2}-\rho_{1}\right)  }\right]
\lambda_{\mathrm{Y}}\right)  \;,\nonumber\\
&
\end{align}
hence%
\begin{equation}
\frac{j_{\mathrm{X}}}{\gamma_{\mathrm{E}}}=q_{\mathrm{d}}^{2}+4\upsilon
q_{12}^{\prime\prime2}+\sum_{n=1}^{2}\sum_{m\geq3}G_{nm}\left\vert
q_{nm}\right\vert ^{2}\;, \label{j_X}%
\end{equation}
and%
\begin{equation}
\frac{j_{\mathrm{Y}}}{\gamma_{\mathrm{E}}}=q_{\mathrm{d}}^{2}+4\upsilon
q_{12}^{\prime2}+\sum_{n=1}^{2}\sum_{m\geq3}G_{nm}\left\vert q_{nm}\right\vert
^{2}\;, \label{j_Y}%
\end{equation}
where $q_{\mathrm{d}}=q_{11}-q_{22}$,%
\begin{equation}
\upsilon=1-\frac{\left(  \mathcal{F}_{11}+\mathcal{F}_{22}-2\mathcal{F}%
_{12}\right)  e_{12}}{2\left(  \rho_{1}-\rho_{2}\right)  }\;,
\end{equation}
$q_{12}^{\prime}=\operatorname*{Re}q_{12}$, $q_{12}^{\prime\prime
}=\operatorname{Im}q_{12}$, and where%
\begin{equation}
G_{nm}=1+\frac{\left(  \mathcal{F}_{1m}-\mathcal{F}_{2m}\right)  e_{nm}}%
{\rho_{1}-\rho_{2}}\;.
\end{equation}
With the help of Eqs. (\ref{rho TE}), (\ref{F(x,y) V2}) and (\ref{f_D}) one
finds that [note that $e_{nm}=-\log\left(  \rho_{n}/\rho_{m}\right)
=\log\left(  \left(  1-\eta_{nm}\right)  /\left(  1+\eta_{nm}\right)  \right)
=-2\eta_{nm}/f_{\mathrm{D}}\left(  \eta_{nm}\right)  $]%
\begin{equation}
\upsilon=\frac{1}{f_{\mathrm{D}}\left(  \eta_{12}\right)  }\;,
\end{equation}
and that $G_{1m}=G\left(  \rho_{1}/\rho_{m},\rho_{2}/\rho_{m}\right)  $ and
$G_{2m}=G\left(  \rho_{2}/\rho_{m},\rho_{1}/\rho_{m}\right)  $, where the
function $G$ is given by%
\begin{equation}
G\left(  r_{1},r_{2}\right)  =1-\frac{\frac{r_{1}-1}{\log r_{1}}-\frac
{r_{2}-1}{\log r_{2}}}{r_{1}-r_{2}}\log r_{1}\;,
\end{equation}
or%
\begin{equation}
G\left(  r_{1},r_{2}\right)  =\frac{r_{2}-1}{r_{2}\log r_{2}}\frac{\log
\frac{r_{1}}{r_{2}}}{\frac{r_{1}}{r_{2}}-1}\;, \label{G(r1,r2)}%
\end{equation}
hence $\upsilon\geq1$ [since $0\leq f_{\mathrm{D}}\left(  \eta_{12}\right)
\leq1$] and $G_{nm}\geq0$ [see Eq. (\ref{G(r1,r2)}), and note that for
non-negative $r_{1}$ and $r_{2}$, both the first factor, which depends on
$r_{2}$ only, and the second one, which depends on $r_{1}/r_{2}$ only, are
non-negative], and thus both $j_{\mathrm{X}}$ and $j_{\mathrm{Y}}$\ are non-negative.

In summary, the dynamics governed by the GME (\ref{TME}) in the vicinity of
the steady state $\rho_{0}$ depends on the $d_{\mathrm{H}}^{2}-1$ diagonal
element of the Jacobean matrix $J_{\mathrm{A}}+J_{\mathrm{B}}$. Our derived
expressions for the eigenvalues, given by Eqs. (\ref{j_l}), (\ref{j_X}) and
(\ref{j_Y}), can be used to evaluate statistical properties of the system near
its steady state $\rho_{0}$. We find that all these eigenvalues are
non-negative, and conclude that the steady state $\rho_{0}$ is stable. This
raises the question under what conditions dynamical instability is possible in
a quantum Hilbert space of finite dimensionality.

We thank Mark Dykman for useful discussions. This work is supported by the
Israel science foundation and by the Israeli ministry of science.

\bibliographystyle{ieeepes}
\bibliography{acompat,Eyal_Bib}

\newif\ifabfull\abfulltrue
\begin{thebibliography}{10}

\bibitem{Fernengel_385701}
Bernd Fernengel and Barbara Drossel,
\newblock ``Bifurcations and chaos in nonlinear lindblad equations'',
\newblock {\em Journal of Physics A: Mathematical and Theoretical}, vol. 53,
  no. 38, pp. 385701, 2020.

\bibitem{Lindblad_119}
Goran Lindblad,
\newblock ``On the generators of quantum dynamical semigroups'',
\newblock {\em Communications in Mathematical Physics}, vol. 48, no. 2, pp.
  119--130, 1976.

\bibitem{Yukalov_9232}
VI~Yukalov,
\newblock ``Nonlinear spin dynamics in nuclear magnets'',
\newblock {\em Physical Review B}, vol. 53, no. 14, pp. 9232, 1996.

\bibitem{Prataviera_01}
GA~Prataviera and SS~Mizrahi,
\newblock ``Many-particle sudarshan-lindblad equation: mean-field
  approximation, nonlinearity and dissipation in a spin system'',
\newblock {\em Revista Brasileira de Ensino de F{\'\i}sica}, vol. 36, no. 4,
  pp. 01--11, 2014.

\bibitem{breuer2002theory}
Heinz-Peter Breuer, Francesco Petruccione, et~al.,
\newblock {\em The theory of open quantum systems},
\newblock Oxford University Press on Demand, 2002.

\bibitem{Drossel_217}
Barbara Drossel,
\newblock ``What condensed matter physics and statistical physics teach us
  about the limits of unitary time evolution'',
\newblock {\em Quantum Studies: Mathematics and Foundations}, vol. 7, no. 2,
  pp. 217--231, 2020.

\bibitem{Hicke_024401}
C~Hicke and MI~Dykman,
\newblock ``Classical dynamics of resonantly modulated large-spin systems'',
\newblock {\em Physical Review B}, vol. 78, no. 2, pp. 024401, 2008.

\bibitem{Levi_053516}
Roei Levi, Sergei Masis, and Eyal Buks,
\newblock ``Instability in the hartmann-hahn double resonance'',
\newblock {\em Phys. Rev. A}, vol. 102, pp. 053516, Nov 2020.

\bibitem{Holstein_1098}
T~Holstein and Hl~Primakoff,
\newblock ``Field dependence of the intrinsic domain magnetization of a
  ferromagnet'',
\newblock {\em Physical Review}, vol. 58, no. 12, pp. 1098, 1940.

\bibitem{Schlomann_S386}
E~Schl{\"o}mann, JJ~Green, and uU~Milano,
\newblock ``Recent developments in ferromagnetic resonance at high power
  levels'',
\newblock {\em Journal of Applied Physics}, vol. 31, no. 5, pp. S386--S395,
  1960.

\bibitem{Grabert_161}
H~Grabert,
\newblock ``Nonlinear relaxation and fluctuations of damped quantum systems'',
\newblock {\em Zeitschrift f{\"u}r Physik B Condensed Matter}, vol. 49, no. 2,
  pp. 161--172, 1982.

\bibitem{Ottinger_052119}
Hans~Christian {\"O}ttinger,
\newblock ``Nonlinear thermodynamic quantum master equation: Properties and
  examples'',
\newblock {\em Physical Review A}, vol. 82, no. 5, pp. 052119, 2010.

\bibitem{Ottinger_10006}
Hans~Christian {\"O}ttinger,
\newblock ``The geometry and thermodynamics of dissipative quantum systems'',
\newblock {\em EPL (Europhysics Letters)}, vol. 94, no. 1, pp. 10006, 2011.

\bibitem{Taj_062128}
David Taj and Hans~Christian {\"O}ttinger,
\newblock ``Natural approach to quantum dissipation'',
\newblock {\em Physical Review A}, vol. 92, no. 6, pp. 062128, 2015.

\bibitem{Bassi_055027}
Angelo Bassi and Kasra Hejazi,
\newblock ``No-faster-than-light-signaling implies linear evolution. a
  re-derivation'',
\newblock {\em European Journal of Physics}, vol. 36, no. 5, pp. 055027, 2015.

\bibitem{kubo2012statistical}
Ryogo Kubo, Morikazu Toda, and Natsuki Hashitsume,
\newblock {\em Statistical physics II: nonequilibrium statistical mechanics},
  vol.~31,
\newblock Springer Science \& Business Media, 2012.

\bibitem{Spohn_33}
Herbert Spohn,
\newblock ``An algebraic condition for the approach to equilibrium of an open
  n-level system'',
\newblock {\em Letters in Mathematical Physics}, vol. 2, no. 1, pp. 33--38,
  1977.

\end{thebibliography}

\end{document}